# A pseudo-simulation of Shor's quantum factoring algorithm

J.F.Schneiderman, M.E.Stanley and P.K.Aravind
Physics Department
Worcester Polytechnic Institute
Worcester, MA 01609

ABSTRACT

This paper presents a computer program, written in Maple, that allows a user to simulate certain aspects of Shor's quantum factoring algorithm on a desktop or laptop computer. The program does not simulate the unitary operations carried out by a quantum computer but does faithfully mimic its output at the crucial "readout" step of the order-finding process. The program requires only two inputs from the user – the number to be factored (which can be up to 10 digits long) and the number of qubits to be used in the factoring (for which a helpful hint is given). The program then returns a detailed history of all its attempts at factoring the number, beginning with its various unsuccessful attempts (together with the reasons for failure) and ending with the final successful attempt that leads to the correct factors. The structure of the simulation is described, a typical output produced by it is shown, and the factors limiting its performance are discussed. The purpose of this simulation is to provide the user with some "hands-on" experience of how quantum factoring works on integers somewhat larger than can be handled by today's quantum computers.

# 1. Introduction

The discovery by Peter Shor[1], in 1994, of an efficient algorithm for factoring large integers aroused a great deal of interest among both scientists and the lay public and also heralded the birth of the new field of quantum computation[2-4]. Shor's discovery was exciting because it used a novel approach, based on quantum mechanics, to greatly speeden up what had been considered a very time consuming task, and it was also very significant because it undermined the security of one of the most widely used cryptographic protocols, the so-called RSA (or Rivest-Shamir-Adelman) scheme. Efforts to build a rudimentary quantum computer capable of running Shor's algorithm got under way almost as soon as Shor had proposed his idea. These efforts recently culminated in the operation of a 7-qubit quantum computer by an IBM-Stanford team[5] that successfully factored the integer 15. Although this may not amount to much in practical terms, it was an experimental feat of heroic proportions and a very significant milestone on the arduous road leading from Shor's vision to reality. While it is not clear how far, or how quickly, one will proceed down this road, the potential payoff is so enormous that there appears to be a great deal of enthusiasm for pressing on with the quest.

Several tutorial accounts of Shor's algorithm are now available in the literature[6-10], mostly at the level of a graduate student who has taken a first course in quantum mechanics. However the student who has read these accounts, and even one who has not, may long for a more direct experience of how a quantum computer operates. It is with the intention of catering to this desire that we have created the simulation presented in this paper. Our simulation, written in Maple and intended to be run on a desktop or laptop computer, tries to provide the user with some feel for how a quantum computer goes about factoring a modestly large integer (of up to about 10 digits) given to it.

Before explaining what our simulation does, and what it does not do, we say a few words about quantum computation in general. A quantum computer operates by storing information in two-state quantum systems (or "qubits"). The action of the computer consists of performing unitary transformations on the qubits and then making a measurement on them from which it attempts to extract the answer to the problem of interest. The unitary transformations are realized physically by "quantum circuits" built up out of a small number of elementary quantum gates that are the analogs of the classical Boolean gates. The final measurement on the qubits collapses their state into one of the so-called "computational basis states", from which the answer to the problem can be extracted with a relatively high probability. A failure at this point is not catastrophic because it can always be detected immediately (as when the incorrect factors of a number are returned), and the calculation repeated, as often as necessary, until the correct answer is obtained (and confirmed).

We can now describe the nature of our simulation more precisely: it avoids modeling all the unitary operations carried out by a quantum computer and merely mimics what it does at the climactic measurement (or "readout") step. Theoretical considerations allow one to predict the probabilities of the various outcomes at the readout step, thereby providing the basis for a simulation of this kind. However the prediction of the probabilities does involve some cheating in that it requires a foreknowledge of the quantity one is trying to calculate (namely, the "order" of a number with respect to another one) before it can proceed. Thus our simulation is in the somewhat embarrassing position of making use of the very result it is after in order to simulate how a quantum computer would go about calculating that result! However this dishonesty (or circularity) of approach seems to be unavoidable if one wishes to simulate the factoring of more than the most trivial numbers. The truth of Feynman's observation[3], that it is impossible to simulate the action of a quantum computer using a classical one, is something that we have come to appreciate much more keenly in the course of working on this project.

We wrote our simulation in the computer algebra program Maple [11] because of its extensive number theory package with standard commands for operations such as the Euclidean algorithm, order finding and continued fraction manipulations, all of which were essential for this project. Another bonus of Maple is its ability to store, and perform calculations with, extremely large integers, which is an advantage if one wishes to simulate the factoring of large numbers. We found, in experimenting with our simulation, that we were limited to factoring integers of about 10 digits not because we had strained the capacities of Maple but because we ran into time or memory limitations on our computer. As the capabilities of laptops and desktops grow, the size of the numbers that can be factored with a simulation of this sort can be expected to creep up somewhat.

Our simulation, the first few lines of which are shown in Table 1, requires the user to supply only two inputs: the number to be factored ($=N$) and the number of qubits in the work register of the quantum computer ($=L$). $N$ can be any composite number of up to 10 digits and should preferably be the product of just two primes, although other possibilities will also work. After the user enters a value of $N$, the program returns a "safe" value of $L$ (i.e. one that guarantees that the factoring will work) and asks the user to enter a value of $L$. The user can simply echo the safe value of $L$, or else enter a different value if he/she wishes. Choosing a smaller value of $L$ can have a variety of effects: it could speeden up the simulation, or it could slow it down, or it could even lead to a failure to obtain any factors at all (if the value of $L$ chosen is too small). After the user has entered the two inputs, the program sets about its work and eventually returns the two factors (or perhaps just some two factors) of the given number, together with a detailed history of all its attempts at the factoring up to the final successful attempt. A sample output produced by our program is shown in Table 3 and will be discussed further in Sec.4.

In addition to the considerable range over which the two user inputs can be varied, our program acquires an extra dimension of richness from the fact that reruns of it with the same user inputs generally lead to quite different factoring histories (that, of course, all converge upon the same correct result). The reason for this plethora of histories is that the program has to make random choices at several points during its execution, and varying choices at these points lead to a "garden of forking paths (universes?)". Our simulation thus not only tells many different stories, but also supplies many alternative plotlines (all leading to the same happy ending) for each story. The user thus has access to a fairly rich variety of scenarios in which to witness the twists and turns of the factoring algorithm as it works its way towards its successful conclusion.

We digress momentarily from our work to mention a couple of other (and more honest) simulations of quantum factoring. A group at Los Alamos[12] carried out a detailed simulation of an ion trap quantum computer factoring the integer 15. The computer they simulated consisted of 18 ions (qubits) that were subjected to nearly 15000 laser pulses that implemented the needed unitary transformations on them. The focus of the study was to see how inaccuracies in the laser pulses, and hence in the unitary transformations induced by them, would affect the performance of the computer and also to study the extent to which "watchdog" stabilization could be used to compensate for these errors. The net conclusion was that error correction techniques were absolutely essential in carrying out even a modest computation of this sort on an ion trap quantum computer. The IBM-Stanford team[5] that operated a 7-qubit quantum computer also simulated the factoring of 15, but focussed on the inclusion of decoherence effects arising from the interaction of the qubits with their environment. Their simulations agreed well with their spectra and confirmed that although decoherence was definitely present in this case, it was not serious enough to derail the process.

This paper is organized as follows. Section 2 describes a simple algorithm for factoring a number and points out that the bottleneck in it, when applied to large numbers, is the "order-finding" step. Section 3 describes Shor's quantum algorithm for determining the order of a number which, when embedded into the algorithm of Sec.2, provides an efficient approach to the problem of factoring very large numbers. Our discussion of Shor's order finding algorithm is a compressed version of the account given by Ekert and Jozsa[6] and serves mainly as a backdrop for the introduction of the principal formulae on which our simulation is based. It is not intended as a first introduction to Shor's algorithm, for which the reader should consult either the Ekert-Jozsa paper or one of the other sources cited.[7-10]  Section 4 discusses some of the problems we faced in creating our simulation and how we resolved them. It also shows a sample output produced by the simulation and comments on some of its features. Finally, the performance of the program in factoring an eight digit number under a fairly wide range of conditions is summarized, in an attempt to convey some feeling for practicalities. Section 5 ends with some brief concluding remarks.

A copy of our Maple program has not been included with this paper to save on space. Arrangements will be made to post it at a suitable site from which it can be retrieved by interested parties. Pending such a posting, a copy of the Maple worksheet of the program (9 kB) can be obtained by emailing a request to paravind@wpi.edu.

**2. A simple algorithm for factoring a number**

Table 1 shows a flowchart of a simple algorithm for determining the factors of a number. Shown with the flowchart is an example of how the algorithm works for a particular number for which there is no need to loop back to an earlier point in the flowchart. The algorithm is very straightforward, but it runs into a bottleneck when applied to very large numbers: the order-finding step in it becomes extremely slow. Several other algorithms for factoring numbers are also known[13], but they all share the feature that the number of elementary operations required for the factoring grows exponentially with the size of the number to be factored. Shor showed how to get around this bottleneck by proposing an efficient quantum algorithm for finding the order of a number that is exponentially faster than the best known classical algorithm. The use of Shor's quantum order finding algorithm within the larger "classical" algorithm of Table 1 leads to a very efficient scheme for factoring extremely large integers.

# 3. Shor's quantum order-finding algorithm

Shor's algorithm works by storing and manipulating information in two distinct sets (or "registers") of qubits: a work register of $L$ qubits and an auxiliary register of $L'$ qubits. The work register is so called because it is where the work of finding the order is done, while the auxiliary register gets its name from the fact that it is used as a sort of scratchpad during the calculation. If $N$ is the number to be factored, the sizes of the work and auxiliary registers are chosen as the integers satisfying the inequalities $N^2 \leq 2^L \leq 2N^2$ and $2^{L'-1} < N \leq 2^{L'}$. The reason for these choices is that the auxiliary register has to be big enough to store all integers from 0 to $N-1$, while the work register must have sufficient room in it to determine the order. The number $q = 2^L$ is the number of "computational basis states" or "readout values" in the work register. We now introduce a compact notation for computational basis states that will greatly simplify the description of the work and auxiliary registers in the account below.

Consider a register of $n$ qubits. Let $|0\rangle$ and $|1\rangle$ denote orthogonal basis states of a single qubit and let $|x_n x_{n-1}...x_1\rangle$ be a state of the register in which the $i$-th qubit is in the state $|x_i\rangle$, where each $x_i$ can be either 0 or 1. We introduce the abbreviated notation $|j\rangle$ for the state $|x_n x_{n-1}...x_1\rangle$, where $j = x_n 2^{n-1} + x_{n-1} 2^{n-2} + ... + x_1 2^0$ is the decimal number whose binary expansion is $x_n x_{n-1}...x_1$. With this notation, an arbitrary state of the work and auxiliary registers can be represented by a superposition of states of the form $|j\rangle|k\rangle$, where the first and second kets refer to computational basis states of the work and auxiliary registers and $j$ and $k$ are decimal numbers lying in the ranges $0 \leq j \leq 2^L - 1$ and $0 \leq k \leq 2^{L'} - 1$.

We now outline the steps of Shor's algorithm for determining the order, $r$, of an integer $y$ with respect to the coprime integer $N$, numbering the steps for convenience.

**Step 0**: Prepare the work and auxiliary registers in the state $|0\rangle|0\rangle$.

**Step 1**: Apply a Hadamard rotation (i.e. the unitary transformation $|0\rangle \to (|0\rangle + |1\rangle)/\sqrt{2}$, $|1\rangle \to (|0\rangle - |1\rangle)/\sqrt{2}$) to each qubit in the work register to obtain the state $\frac{1}{\sqrt{q}} \sum_{a=0}^{q-1} |a\rangle|0\rangle$, which represents a uniform superposition of all the computational basis states in the work register.

**Step 2**: For each number $a$ in the work register, calculate the quantity $y^a \mod N$ and store the result in the auxiliary register. This produces the following entangled state of the work and auxiliary registers:
$$\frac{1}{\sqrt{q}} \sum_{a=0}^{q-1} |a\rangle |y^a \mod N\rangle \qquad (1)$$
This step is carried out in a quantum computer by a quantum circuit that performs the modular exponentiation operation efficiently. It is worth stressing that a single pass of the state in Step 1 through the quantum circuit suffices to calculate the function $y^a \mod N$ for all values of $a$, courtesy of the superposition principle of quantum mechanics. This is where some quantum magic occurs, because a classical computer would be unable to perform a massively parallel calculation like this without an enormous amount of additional hardware.

**Step 3**: Make a measurement on the auxiliary register in the computational basis. Suppose this yields the state $|z\rangle$, where $l$ is the smallest integer such that $z \equiv y^l \mod N$. Then the state of the work register collapses to

$$|\Phi_l\rangle = \frac{1}{\sqrt{A+1}} \sum_{j=0}^{A} |jr+l\rangle \qquad (2)$$

where $A$ is the largest integer less than $(q-l)/r$. We will take $A = q/r - 1$ henceforth, this being the exact value of $A$ if $r$ is a factor of $q$ and differing from the exact value by an insignificant amount otherwise. Note that the only states surviving in the work register are separated from each other by the sought for order $r$. However a single measurement on this state (which is all we are allowed!) will fail to reveal the order because of the unknown offset $l$. It is at this point that Shor produced his masterstroke.

**Step 4**: Shor proposed applying a quantum Fourier transform to the state (2) to get

$$\text{QFT}_q |\Phi_l\rangle = \sum_c \tilde{f}(c) |c\rangle \qquad \text{where} \qquad \tilde{f}(c) = \frac{\sqrt{r}}{q} \exp\left(2\pi i \frac{lc}{q}\right) \sum_{j=0}^{q/r-1} \exp\left(2\pi i \frac{jrc}{q}\right) \qquad (3)$$

The Fourier transform converts the offset $l$ into a harmless phase factor and the original function of period $r$ into a new function of period $q/r$. A single measurement on the transformed work register serves to yield the new period, and hence the order, with a relatively high probability, as we will see below. Quantum circuits that perform the quantum Fourier transform efficiently are known.[8] This step is perhaps the single most important step in the entire algorithm, and the one on which its success crucially hinges.

**Step 5**: Carry out a measurement on the work register in the computational basis. This yields the state $|c\rangle$ with probability

$$\text{Prob}(c) = |\tilde{f}(c)|^2 = \frac{r}{q^2} \left| \sum_{j=0}^{q/r-1} \exp\left(2\pi i j \frac{rc \mod q}{q}\right) \right|^2. \qquad (4)$$

Note that this probability is independent of either $z$ or $l$, the quantities determined by the measurement on the auxiliary register in Step 3. This may give some feeling for why Step 3 is unnecessary for the order finding process and could have been omitted (the main reason for including this step in the discussion is that it serves as a convenient mental crutch, as becomes obvious if one tries to dispense with it).

**Step 6**: The next step is to calculate the probabilities in (4) as a basis for our simulation. We begin by noting, from Fig.1, that for any $c$ there is exactly one integer, which we denote $m_c$, such that $-q/2 \leq rc - m_c q \leq q/2$. Introduce the angle $\theta_c = (2\pi/q)(rc - m_c q)$, which lies between $-\pi$ and $\pi$. The geometrical meaning of $\theta_c$ is that it is the angle between neighboring phasors in the sum in (4). The resultant of all the phasors in (4) can be found by summing a geometric series and, on taking the squared length of this resultant, one finds that the probability of obtaining the "readout" value $c$ upon making a measurement on the work register is

$$\text{Prob}(c) = \frac{r}{q^2} \frac{\sin^2\left(\frac{\theta_c q}{2r}\right)}{\sin^2\left(\frac{\theta_c}{2}\right)} \quad \text{where} \quad \theta_c = 2\pi \frac{(rc - m_c q)}{q}. \quad (5)$$

For the vast majority of values of $c$, the angle $\theta_c$ has an appreciable value and the phasors in (4) loop around a circle of small radius numerous times to give a practically vanishing resultant (and hence probability). However, for a relatively small number of $c$ values, $\theta_c$ is vanishingly small and the phasors practically line up with each other to give rise to a large resultant, and hence probability. In particular, the $r$ values of $c$ satisfying the inequalities $-r/2 \leq rc - mq \leq r/2$ (for suitable $m$) give rise to phasors that curve around no more than a (large) semicircle at most, and hence lead to appreciable probabilities. To summarize, the $r$ values of $c$ just mentioned, together with the $c$ values in their immediate neighborhood, give rise to the "dominant" or non-negligible probabilities, while all other $c$ values give rise to practically vanishing probabilities. Our simulation uses a random number generator to pick readout values in accordance with their probabilities, as specified by (5). More details of this will be provided in Sec.4.

**Step 7**: The only task that now remains is to extract the order, $r$, from the "readout" value of $c$ obtained in Step 6. To do this, we recast the inequalities for the "dominant" $c$ values given a few lines earlier in the form

$$\left|\frac{c_m}{q} - \frac{m}{r}\right| < \frac{1}{2q} < \frac{1}{2N^2} < \frac{1}{2r^2}, \quad (6)$$

where the last two inequalities follow from the fact that $q \geq N^2$ and $N \geq r$. The inequality (6) guarantees, from a theorem in number theory, that the quantity $m/r$ can be found as the highest convergent of the continued fraction expansion of $c_m/q$ whose denominator is less than $N$. This procedure yields $m/r$ and hence $r$ if $m$ and $r$ are coprime, but it fails if $m$ and $r$ have any common factors. One can always tell if the correct value of $r$ has been obtained (by calculating $y^r \bmod N$ and seeing if it is equal to 1) and, if it has not, return to Step 0 and repeat the entire process over again, as often as necessary, until the correct order is obtained and confirmed.

How often does the above process have to be repeated, on average, before the correct order is obtained? Ekert and Jozsa used some elementary number theoretic considerations to show that the average number of repetitions is less than $\log N$ for large $N$ However, for the relatively small values of $N$ involved in our simulations, we found that success was always achieved in far fewer attempts than predicted by this (rather weak) asymptotic estimate.

## 4. Details of the simulation

To illustrate some of the problems we ran into in creating our simulation, and how we resolved them, consider the problem of factoring a relatively small number like $N = 187$ using $L = 16$ qubits in the work register (which is the "safe" number of qubits needed to perform this factoring). The work register then has a total of $q = 2^{16} = 65536$ readout values in it. Suppose first that $y = 56$ is chosen as the number whose order has to be determined. The order in this case is $r = 16$, which is a factor of $q = 65536$, and (5) then implies the simple scenario (mentioned in most accounts of quantum factoring) of $r$ uniformly spaced readout values with a probability of $1/r$ each and all other readouts having a probability of zero (see Fig.2(a) for a plot of these probabilities in the present case). The probabilities in any case like the present one, in which $r$ is a factor of $q$, are trivial and the simulation poses no problems.

However matters become rather more complicated if the order to be determined is not a factor of $q$. To illustrate the difficulties that arise in this (far more typical) case, suppose that $y = 36$ had been chosen in the above case instead as the number whose order had to be determined. Then the order ($r = 40$) is not a factor of $q$ and a plot of the probabilities of the readout values, shown in Fig.2(b), reveals a more complex structure than the simple periodic pattern shown in Fig.2(a). A careful inspection of Fig.2(b) reveals the following features: (1) there are still $r$ (= 40) "dominant" readout values whose probabilites dominate those of the rest; (2) the dominant probabilities are no longer equal to each other but consist of eight repetitions of a basic set of five (because $40 = 2^3 \cdot 5$); (3) the sum of the dominant probabilities is .7792, which is significantly less than 1; (4) the non-dominant readouts in the immediate vicinity of the dominant ones have non-negligible probabilities associated with them (this may be seen from the dots halfway down some of the peaks); and (5) the vast majority of readouts still continue to have almost vanishing probabilities associated with them.

Points (2),(3) and (4) highlight the difficulties of carrying out a realistic simulation: it is the task of calculating all the nontrivial probabilities involved. There were two strategies we adopted to get around this obstacle. The first was a standard programming trick that helped cut down on the calculational load, while the second was a somewhat biased selection of one of the random inputs needed by the program. We discuss each of these strategies in turn.

The programming trick we used was that of calculating only the probabilities needed up to any point in the simulation, and no more. This works in detail as follows. The interval from 0 to 1 is divided into bins corresponding to the readout values, with each bin having a width equal to the probability of its readout value. Our simulation picks a readout value by asking a random number generator to return a number between 0 and 1 and then picking the bin into which that number falls. The crucial trick is to construct the bins only after the random number has been returned, since one can then get away with constructing the minimum number of bins (in fact, just as many needed to trap the random number returned). The trick is further enhanced if one constructs the bins in order of decreasing probability rather than increasing readout value i.e. one first constructs the bins corresponding to all the dominant readouts, then the bins corresponding to the readouts in the vicinity of the dominant readouts, and finally the bins corresponding to the vast majority of unimportant readouts. All bins constructed during a particular subcycle of a simulation are stored, so that they do not have to be recomputed if they are needed again. This approach provided a simple way of getting around the obviously impractical task of calculating all the probabilities involved.

However the above strategy alone did not suffice when it came to factoring 10 digit numbers. The reason is that the number of dominant readouts in such a case could be in excess of a million, leading to very time consuming calculations even when a relatively small random number is returned. Our way of getting around this obstacle is to simply have our program reject any randomly chosen $y$ whose order exceeds the square root of the number being factored. This automatically places a ceiling on the number of dominant readouts (which is equal to the order) and prevents the calculation of the probabilities from stretching to unacceptably long times. There is of course some distortion involved in this step, in that it makes our simulation a little less authentic, but we felt that it was a worthwhile price to pay in order to extend the reach of our program upwards to larger integers. The user who wishes to disable this feature and perform a truly honest simulation, at the cost of waiting a much longer time, can easily do so.[14]

Having discussed the hurdles that had to be overcome in devising our simulation, we now discuss the sort of output it produces. Table 3 shows the output produced by our program when asked to factor $N = 1328881$ using $L = 41$ qubits in the work register. The output begins by listing $N$ and $L$ and then recounts the details of all the attempts made by the program to factor $N$. Each unsuccessful attempt consists of two steps: selecting a random number $y$ coprime to $N$, and then determining the order of $y$ by a pseudo-simulation of the action of a quantum computer. The order determination step typically involves several "trials", because the initial unsuccessful trials yield readouts that lead, via the continued fraction algorithm, to spurious orders. When the order has been correctly determined, the attempt at factoring could still fail because the order is odd or else because it is the wrong type of even number. After a train of unsuccessful attempts comes a final successful attempt that yields an order leading to nontrivial factors of $N$. The output concludes by listing the factors of $N$ and also mentioning the time and the number of trials (of quantum order-finding) taken to perform the factorization.

The above remarks should suffice to follow a typical output of the program, such as the one shown in Table 3. A few other points should be mentioned. If, as sometimes happens, the randomly chosen value of $y$ is not coprime to $N$, the greatest common denominator of $y$ and $N$ is returned as a factor and the program terminates. If the user should happen to choose a prime number to factor, he/she is warned about this and told to go back and choose another number (ignoring this warning and forging ahead could lead to a long wait!). The program works best if the number to be factored is the product of just two primes. This can always be done by arming onself with a list of primes[15] and multiplying two of the numbers on the list. Entering a number that is the product of more than two primes could lead to factors that are not prime and whose product is not always equal to the number being factored. Choosing a value of $L$ appreciably less than the safe value required for factoring could sometimes cause the program to terminate without finding the factors; we deliberately built this eventuality into our program by limiting the number of order-finding trials to 100, in order to avoid having our program get caught up in an infinite loop.

How long does our simulation take to factor moderately large numbers? To give some feeling for this, we factored $N = 25610987$ using various numbers of qubits $(= L)$ in the work register and tabulated the running time of the simulation as well as the number of trials (of quantum order-finding) in Table 4. Each row of the table shows the time (and number of trials) for four runs carried out with a fixed value of $L$, with the fifth entry being the average time (number of trials) computed over the previous four entries. The first row is for $L = 50$, the "safe" value of $L$, whereas the later rows are for progressively smaller values of $L$. Note that there is a considerable variation in the numbers even within a row because the factoring time depends upon a number of chance variables including, most importantly, the unknown order to be determined. It is interesting that the factorization sometimes succeeds even with an $L$ as low as 30, which is appreciably below the safe value. The factoring time averaged around 5 minutes for this number, which is not intolerable. The calculations leading to Table 4 were performed on a laptop with a 200MHz clock and 80MB of RAM. The version of Maple used was Maple V Release 4. We found that running the simulation on a more powerful laptop loaded with Maple VII yielded vastly improved performance, allowing us to factor up to 10 digit numbers, but we did not compile any detailed statistics in this case.

## 5. Conclusion

The purpose of this program is to allow users to explore some aspects of Shor's algorithm for themselves in an experiential setting. The program may be found most useful by students who have taken a course in quantum mechanics or quantum information that includes some discussion of quantum circuits and algorithms. However it might also be used, with suitable guidance, by a less sophisticated audience. We would welcome hearing from readers who use this program and have any comments or feedback to offer.

This paper is based on a Major Qualifying Project (senior thesis) done at Worcester Polytechnic Institute by J.F.S and M.E.S under the guidance of P.K.A.

15. The Maple command ithprime(n) returns the n-th prime number. Embedding this command within a do loop will generate a list of primes.

TABLE CAPTIONS

Table 1. Flowchart of an algorithm for factoring a number , illustrated by the factorization of 187. The notation $a \equiv b \bmod N$ means that $a$, when divided by $N$, leaves the remainder $b$. The notation gcd(a,b) is used to indicate the greatest common denominator of a and b.

Table 2. The first few lines of our Maple program that simulates quantum factoring.

Table 3. The output produced by our Maple program when asked to factor 1328881 using 41 qubits in the work register of the quantum computer.

Table 4. Factorization of $N = 25610987$ with various numbers of qubits $(= L)$ in the work register. The entries in a particular row pertain to a fixed value of $L$ and show the time, in secs, (and the number of trials required for the factoring); the first four entries are for different runs, while the fifth is the average of the previous four entries. Note: a dash for the number of trials indicates that the program was unable to factor the number in the maximum of 100 trials permitted to it.

FIGURE CAPTIONS

Figure1. The top row shows dots spaced a distance $r$ apart and the bottom row shows crosses spaced a distance $q$ apart. Note that any dot in the top row lies between two crosses in the bottom row and is at a distance of less than $q/2$ from the closer of these crosses, whose distance from the leftmost cross (in units of $q$) is the integer $m_c$ in Step 6 of Shor's algorithm, Sec.3. Note also that any cross in the bottom row lies between two dots in the top row and is at a distance of less than $r/2$ from the closer of these dots, whose distance from the leftmost dot (in units of $r$) is the $c$-value of one of the "dominant" readouts in the work register.

Figure 2. Factoring of $N = 187$ with $L = 16$ qubits. The top figure, (a), shows the probabilities of the $2^{16} = 65536$ readout values in the work register for $y = 56$ (order = 16), while the bottom figure, (b), is a similar plot for $y = 36$ (order = 40). Note that when the order of $y$ is a factor of the total number of readout values, as in (a), the pattern of probabilities is very simple, but that the pattern gets much more complicated when this condition is not satisfied.

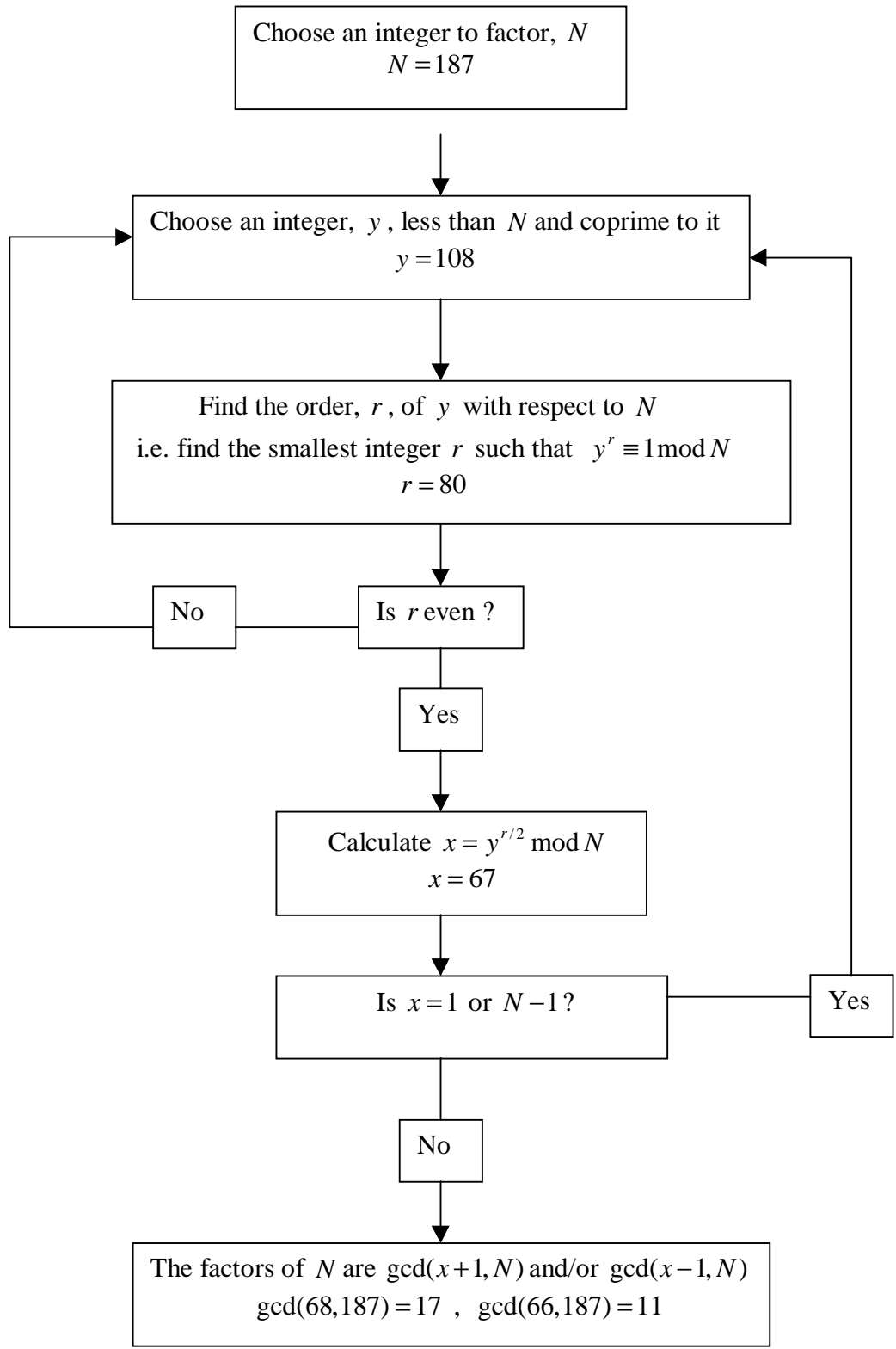

# A SIMULATION OF SHOR'S QUANTUM FACTORING ALGORITHM

User is prompted for two inputs, N and L, at the beginning of the program.

> restart:with (numtheory):with (stats):

Enter a number to factor (=N) below. This number should not be more than 10 digits long.

> N:= 1328881:

> x:=isprime(N):

> if (x=true) then

> lprint(cat(`THE NUMBER YOU PICKED IS PRIME, PLEASE TRY AGAIN!!!`)):

> else

> lprint(cat(`The number to be factored is `, N, `.`)):

> l:= ceil( log[2](N^2) ):

> lprint(cat(`The safe number of qubits needed to factor this number is `, l, `.`)):

> fi:

The number to be factored is 1328881.

The safe number of qubits needed to factor this number is 41.

Enter the number of qubits to be used in the factoring (= L) below. You may simply enter the safe value above if you wish.

> L:=41:

Now hit enter and the program will proceed to factor N.

TABLE 3

The number to be factored is N=1328881.
The number of qubits used to do the factoring is L=41.
-------------------------HISTORY OF FACTORIZATION-------------------------
--------------------------------------------------------------------------
--------------------------------------------------------------------------
Finding order of y = 171891.
--------------------------------------------------------------------------
Trial #1.
The readout value from the work register is 29659273196.
The order found using this readout value is 519.
The order is incorrect, the quantum computer will be reset to try again.
--------------------------------------------------------------------------
Trial #2.
The readout value from the work register is 656741049346.
The order found using this readout value is 519.
The order is incorrect, the quantum computer will be reset to try again.
--------------------------------------------------------------------------
Trial #3.
The readout value from the work register is 1495674776898.
The order found using this readout value is 519.
The order is incorrect, the quantum computer will be reset to try again.
--------------------------------------------------------------------------
Trial #4.
The readout value from the work register is 1794386028375.
The order found using this readout value is 1038.
The quantum computer has found the correct order.
The factors of 1328881 are determined to be 1328881 and 1.
The factoring has failed, hence a new value of y will be chosen.
--------------------------------------------------------------------------
--------------------------------------------------------------------------
Finding order of y = 1328740.
--------------------------------------------------------------------------
Trial #5.
The readout value from the work register is 753655857537.
The order found using this readout value is 213.
The quantum computer has found the correct order.
The order is odd, hence a new value of y will be chosen.
--------------------------------------------------------------------------
--------------------------------------------------------------------------

TABLE 3 (cont'd)

Finding order of y = 505980.
-------------------------------------------------------------------------
Trial #6.
The readout value from the work register is 1671511896561.
The order found using this readout value is 346.
The order is incorrect, the quantum computer will be reset to try again.
-------------------------------------------------------------------------
Trial #7.
The readout value from the work register is 1366445086543.
The order found using this readout value is 346.
The order is incorrect, the quantum computer will be reset to try again.
-------------------------------------------------------------------------
Trial #8.
The readout value from the work register is 1135526459514.
The order found using this readout value is 519.
The order is incorrect, the quantum computer will be reset to try again.
-------------------------------------------------------------------------
Trial #9.
The readout value from the work register is 2137586189645.
The order found using this readout value is 1038.
The quantum computer has found the correct order.
The factors of 1328881 are determined to be 1328881 and 1.
The factoring has failed, hence a new value of y will be chosen.
-------------------------------------------------------------------------
-------------------------------------------------------------------------
Finding order of y = 200298.
-------------------------------------------------------------------------
Trial #10.
The readout value from the work register is 656741049346.
The order found using this readout value is 519.
The quantum computer has found the correct order.
The order is odd, hence a new value of y will be chosen.
-------------------------------------------------------------------------
-------------------------------------------------------------------------
Finding order of y = 205920.
-------------------------------------------------------------------------
Trial #11.
The readout value from the work register is 1535926647664.
The order found using this readout value is 1038.
The quantum computer has found the correct order.
The factors of 1328881 are determined to be 1039 and 1279.
The program has succeeded and will now terminate.
This simulation took 113.895 seconds and 11 trials to factor 1328881.

| FACTORIZATION OF $N = 25610987$ (= 3623·7069) | | | | | |
|---|---|---|---|---|---|
| $L = 50$ | 423(6) | 35(1) | 113(2) | 566(4) | 284(3.3) |
| $L = 46$ | 103(3) | 528(6) | 374(5) | 169(2) | 294(4) |
| $L = 42$ | 86(6) | 105(1) | 611(7) | 283(3) | 271(4.3) |
| $L = 38$ | 237(1) | 867(8) | 384(2) | 1228(15) | 679(6.5) |
| $L = 34$ | 130(3) | 90(9) | 137(2) | 790(38) | 327(13) |
| $L = 30$ | 395(-) | 872(-) | 266(56) | 478(64) | -- |

| $0r$ | $1r$ | $2r$ | | $cr$ | | $qr$ |
|---|---|---|---|---|---|---|
| • | • | • | • | • | • | • |
| × | × | | × | | × | × |
| $0q$ | $1q$ | | | $mq$ | | $rq$ |